\documentclass[aip,jmp,amsmath,amssymb,reprint]{revtex4-1}

\usepackage{graphicx}
\usepackage{dcolumn}
\usepackage{bm}
\usepackage{color,soul}
\usepackage[mathlines]{lineno}

\usepackage{amsmath}
\usepackage{amssymb}
\usepackage{amsfonts}
\usepackage{mathrsfs}
\usepackage{epsfig}
\usepackage{bm}

\begin{document}

\title{Signatures of the self-modulation instability of relativistic proton bunches in the AWAKE experiment}

\author{M. Moreira}
\affiliation{GoLP/Instituto de Plasmas e Fus\~ao Nuclear Instituto Superior T\'ecnico (IST), Lisbon, Portugal \\ CERN, Geneva, Switzerland}
\author{J. Vieira}
\affiliation{GoLP/Instituto de Plasmas e Fus\~ao Nuclear Instituto Superior T\'ecnico (IST), Lisbon, Portugal}
\author{P. Muggli}
\affiliation{Max Planck Institute for Physics, Munich, Germany, \\ CERN, Geneva, Switzerland}

\begin{abstract}
We investigate numerically the detection of the self-modulation instability in a virtual detector located downstream from the plasma in the context of AWAKE. We show that the density structures, appearing in the temporally resolving virtual detector, map the transverse beam phase space distribution at the plasma exit. As a result, the proton bunch radius that appears to grow along the bunch in the detector results from the divergence increase along the bunch, related with the spatial growth of the self-modulated wakefields. In addition, asymmetric bunch structures in the detector are a result of asymmetries of the bunch divergence, and do not necessarily reflect asymmetric beam density distributions in the plasma.
\end{abstract}

\pacs{Valid PACS appear here}

\maketitle

\section{\label{sec:intro}Introduction}
The AWAKE experiment aims to drive plasma wakefields with a relativistic proton (p$^+$) bunch and to accelerate electrons (e$^-$) in these wakefields~\cite{bib:awake}. The p$^+$ bunch ($\sigma_z=$ 6-12\,cm) is much longer than the period of the wakefields ($\lambda_{pe}=$ 3-1\,mm) in a plasma with sufficient e$^-$ density (n$_e=$ (1-10)$\times$10$^{14}$\,cm$^{-3}$) to sustain GV/m amplitude accelerating fields. The experiment relies on a self-modulation process~\cite{bib:kumar} forming a p$^+$ bunch train with a period equal to that of the wakefields. This train can then resonantly drive the wakefields to large amplitudes. The modulation process is seeded by sudden ionization of a rubidium (Rb) vapor by a femtosecond-long laser pulse traveling within the p$^+$ bunch. This process seeds the self-modulation instability~\cite{bib:muggli}.

The effect of the seeded self-modulation (SSM) process is detected and characterized by three p$^+$ bunch diagnostics~\cite{bib:muggli}: a two-screen method~\cite{bib:turner}, the time resolved emission of the optical transition radiation (OTR) emitted by the protons~\cite{bib:rieger}, and spectral analysis of the coherent transition radiation emitted by the p$^+$ bunch train~\cite{bib:martyanov}.

These diagnostics measure the p$^+$ bunch characteristics after its propagation over a few meters of vacuum after the plasma exit. It is therefore important to understand the effect of the plasma wakefields on the transverse phase space distribution of the proton beam (direction, amplitude and global emittance) to properly interpret the experimental results. For example, it is important to realize that the transverse size of the micro-bunches as observed at the diagnostic is different from that in the plasma. In particular, its variation as observed along the bunch is also likely to be different than that in the plasma. This can have important implications on the effectiveness of the driving of the wakefields in the plasma. It can also give information about the evolution of the wakefields along the p$^+$ bunch.

\section{Method}
To this end, we perform 3D simulations of the SSM process using the code Osiris~\cite{bib:fonseca_book,bib:osiris}. We use a 3D geometry to properly describe the p$^+$ bunch emittance and to properly describe the role of beam break-up instabilities, which can only be correctly described in 3D~\cite{bib:vieira}. For this preliminary study we only simulate the first 21 wakefield periods or micro-bunches in order to decrease the computation time. We choose a relatively low plasma e$^-$ density (n$_e$=2$\times$10$^{14}$\,cm$^{-3}$) corresponding to a wakefield period of  7.8\, ps, since experimentally, time-resolved images have a resolution of $\sim$1\,ps.

While in the experiment seeding is obtained from a relativistic plasma ionization front, in the simulations seeding is obtained from a sharp rising leading edge of the p$^+$ bunch charge distribution. Therefore, the cut p$^+$ bunch propagates in a preformed, constant density plasma. This is again to decrease computation time. We are currently investigating the possible differences between these two seeding methods in the development of the SSM. The effect of the finite plasma radius of the experiment is also neglected in this study. This effect has been investigated earlier for SSM of low energy e$^-$ bunches~\cite{bib:fang} and will be investigated further for the AWAKE case.

The bunch relativistic factor is $\gamma_b = 479$ (corresponding to the proton bunch energy at the Super Proton Synchrotron at CERN). The proton bunch density profile is $(n_b/2)\left[ 1+\cos(\pi x/\sigma_z )\right]\exp\left[ -r^2/(2\sigma_r^2)\right] $ for $L_b<x<0$. Here, the peak bunch density is $n_b = 4.1\times10^{12} \textrm{cm}^{-3}$, corresponding to a normalised peak density $n_b/n_0=2.2\times10^{-2}$, the bunch transverse size is $\sigma_r = 160~\mu\mathrm{m}$ or $ 0.42~c/\omega_p$ and the bunch length is $\sigma_z = 15~\mathrm{cm}$ or $398.8~c/\omega_p$. The bunch length corresponding to 21 plasma wavelengths is $L_b = 4.9~\mathrm{cm}$ or $L_b = 140~c/\omega_p$. Here $\omega_p$ is the plasma e$^-$ angular frequency and $c/\omega_p$ the cold plasma skin depth.

The simulation box is in a moving window traveling at the speed of light with dimensions $140 \times 8 \times 8~(c/\omega_p)^3$ in the longitudinal and transverse directions, respectively. This corresponds to a box with $53.0 \times 3.0 \times 3.0$\, mm$^3$. The simulation box is divided into $1690\times216\times216$ cells. This gives 12 cells per plasma skin depth in the longitudinal direction, and 27 cells per skin depth in the transverse directions. Each cell contains $2\times 1\times 1$ simulation particles per species (i.e. p$^+$ and plasma e$^-$). Plasma ions form an immobile continuous background. This assumption also holds in experiments, and is important to prevent the early saturation of the SSM due to the motion of plasma ions~\cite{bib:vieira_prl,bib:vieira_pop}.

To replicate the actual detection method in experiments, a virtual detector placed at a given distance from the plasma exit is modelled. The detector is a spatial grid located on the plane perpendicular to the propagation direction. The distance of the detector to the plasma exit can be changed. The charge of each proton reaching the detector is deposited in the nearest detector cell. The proton trajectories are calculated using a post-processing algorithm that pushes particles until they cross the detector plane. The proton trajectories in vacuum start to be calculated from their positions and momenta at the plasma exit in the Osiris simulations.

\section{Numerical Results}

\begin{figure}
\includegraphics[width=\columnwidth]{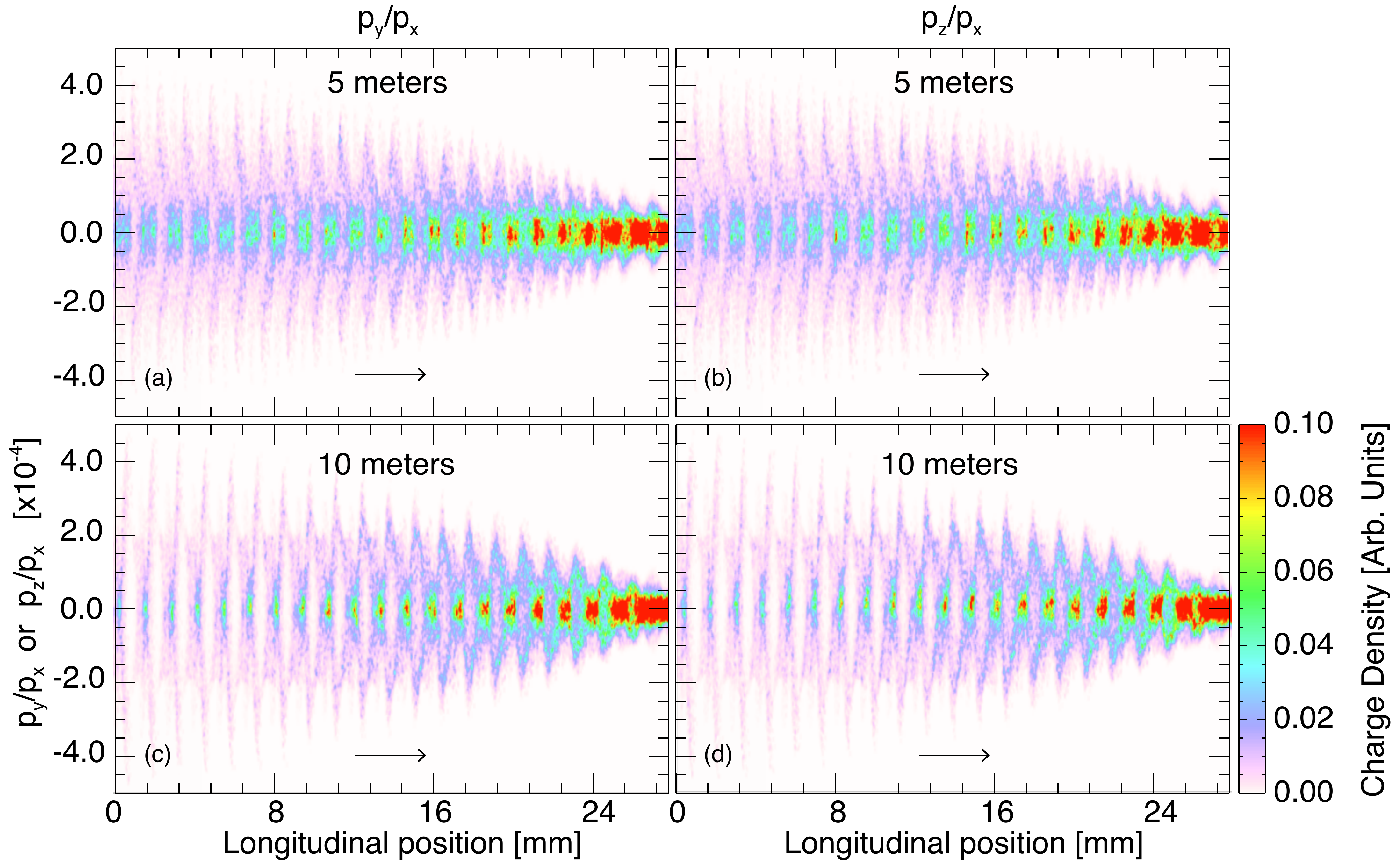}
\caption{Transverse angular distribution of the p$^+$s at two different locations in the plasma and in the two transverse planes. (a) and (b) show the beam divergence in the two transverse planes, $p_y/p_x$ (a) and $p_z/p_x$ (b) after 5\,m in the plasma. (c) and (d) show the beam divergence after 10\,m in the two transverse planes. A fraction of the p$^+$s has left the simulation box for frames (c) and (d). The propagation direction x (from left to right) is indicated by the arrow.}
\label{fig:p2p3}
\end{figure}

Figure~\ref{fig:p2p3} shows the angular distribution of the p$^+$s after propagation over 5\,m (Figure~\ref{fig:p2p3}a-b) and 10\,m (Figure~\ref{fig:p2p3}c-d) of plasma, both in the y and z transverse directions. It is clear that the SSM process has developed and has created the periodic modulation of the p$^+$s momentum with the plasma period. The pattern consists of micro-bunches characterized by a small divergence angle (typical value of $<10^{-4}$\,rad, as seen in Fig.~\ref{fig:p2p3}), though the angle increases somewhat along the train. The periodic pattern does not start immediately at the bunch front and the micro-bunches are getting shorter along the train. The defocused particles located between these micro-bunches have a larger diverging angle, increasing along the train. They also form v-shape features pointing away from the axis, a pattern expected since the wakefields defocusing force is maximum between the micro-bunches. The patterns are similar in the two planes, a feature expected from the development of the SSM, the symmetric version of the modulation mode, which, when effectively seeded, dominates its asymmetric version, the hose instability~\cite{bib:witthum,bib:vieira}.

After the plasma exit, the momentum distribution of the self-modulated proton bunch does not change appreciably. There is no space charge effect in a relativistic beam, as the beam self force scales with $1/\gamma_b^2$, which is negligible for a relativistic beam. This means that p$^+$s propagate along straight paths, which are defined by their velocity at the plasma exit. We also note that here we do not include the potential effect of a plasma ramp at the plasma exit.

\begin{figure}
\includegraphics[width=\columnwidth]{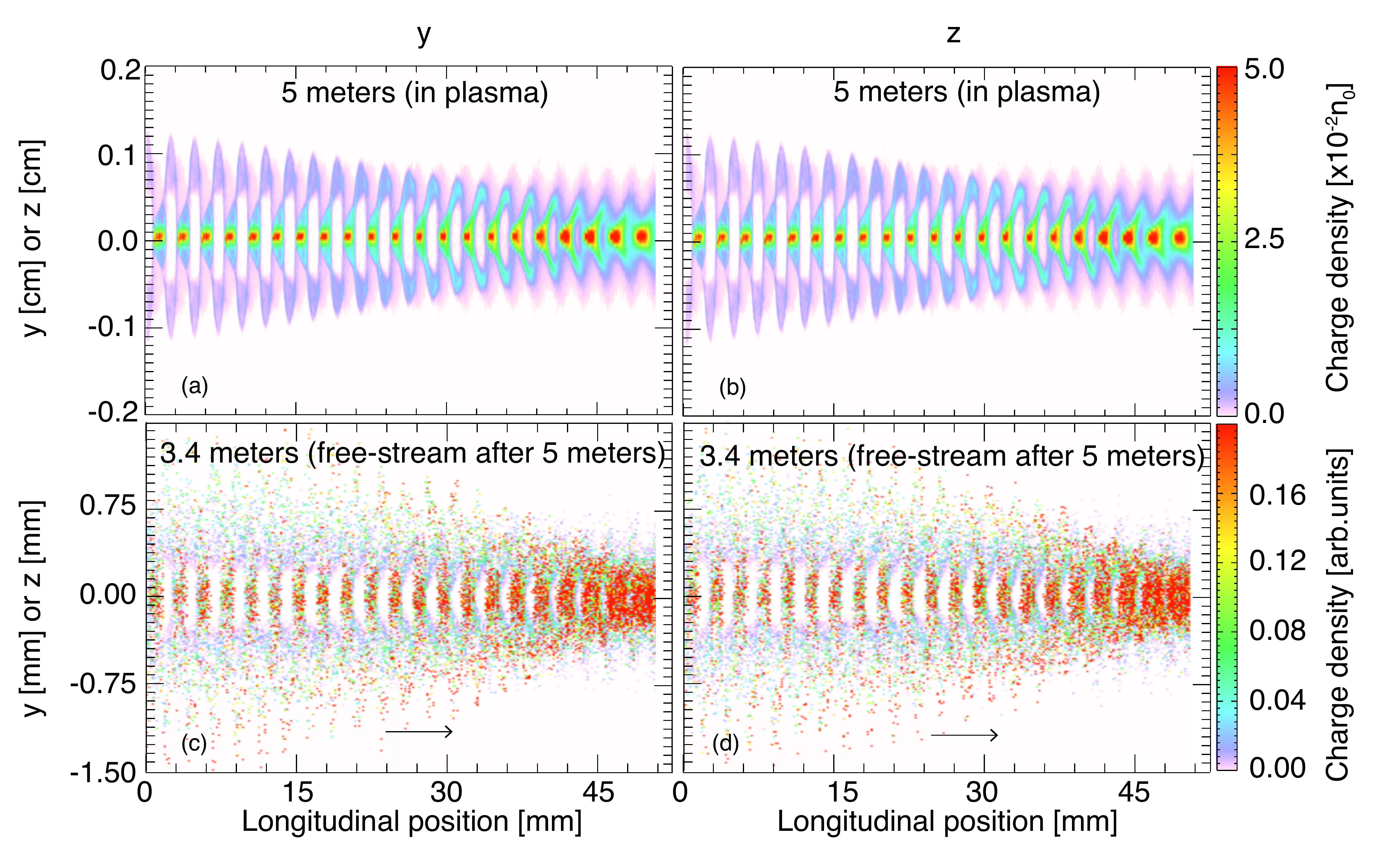}
\caption{Bunch density distribution after 5\,m in the plasma (a)-(b) and at a virtual detector located downstream the plasma (c)-(d). %
The panels (c)-(d) contain all the particles saved by the simulation and that hit a detector centered on the beam axis. The virtual detector is aligned with the axis of the bunch and collects particles within a region which is 4.0\,mm wide along the transverse direction normal to the detector plane. Panels (a)-(b) are a slice of the proton bunch density taken at the center of the simulation box. The thickness of the slice of the simulation is 0.013\,mm.
The arrows indicate the propagation direction of the bunch (from left to right).}
\label{fig:AtDetectorandPlasma}
\end{figure}

Figures~\ref{fig:AtDetectorandPlasma}a-b show the transverse spatial distribution of the p$^+$ just before exiting the plasma, i.e., the bunch pattern that drives the wakefields. The image shows the average proton bunch density along the direction that is normal to the image plane. The thickness of this averaged region is 0.2\,mm, which corresponds to the width of the core of the self-modulated beamlets. We placed a virtual detector to collect p$^+$s 3.4\,m away from the plasma exit. The virtual detector consists of a 2D spatial grid (y-x) or (z-x), where the protons are deposited. Figure~\ref{fig:AtDetectorandPlasma}c-d show the transverse distribution of the p$^+$ after propagation 3.4\,m downstream from the plasma exit.

In order to mimic the functioning of a detector in experiments, the virtual detector is aligned with the axis of the bunch and collects particles within a region which is 4.0\,mm thick along the transverse direction normal to the detector plane. The thickness was chosen in order to include all protons within the core of the self-modulated beamlets. For these figures, the vast majority of the p$^+$ are still inside the simulation box. A small number of defocused p$^+$s have been absorbed at the simulation transverse walls along the plasma. As expected, the patterns are very similar to the momentum patterns of Fig.~\ref{fig:p2p3}.

Comparison between the transverse size of the beam at the plasma exit (given by Fig.~\ref{fig:AtDetectorandPlasma}c-d) and at the detector (given by Fig.~\ref{fig:AtDetectorandPlasma}a-b) reveals that the beam size is typically ten times larger at the detector than at the plasma exit. Thus, the images on the virtual detector mostly reflect the transverse momentum distribution at the plasma exit, rather than the actual beam density at that position. This explains the resemblance between the momentum patterns in Fig.~\ref{fig:p2p3} and the density structures in Fig.~\ref{fig:AtDetectorandPlasma}a-b.

To stress the latter idea further, we note that, although the bunch spatial distribution at the plasma exit is fairly symmetric around $y=0$, the distribution of particles away from the plasma shows the development of asymmetric structures, visible at larger radii. Therefore, the asymmetry observed downstream from the plasma does not reflect the degree of the spatial asymmetry in the plasma, and hence of the degree of hosing. Transverse momentum distribution asymmetries could seed the hosing instability~\cite{bib:schroeder_pre}. In the linear regime, however, the hosing instability does not grow after the saturation of self-modulation~\cite{bib:vieira_prl}.

It is also straightforward to interpret these results from a more quantitative perspective. The position of a beam particle on the detector plane with transverse velocity $\mathbf{v}_{\perp}$ is given by $\mathbf{x}_{\perp} = \mathbf{x}_0+\mathbf{v}_{\perp}d/c$, where $d$ is the propagation distance after the plasma and $\mathbf{x}_0$ the initial position. We can consider that particles are located within $\mathbf{x}_0\lesssim \sigma_r$, where $\sigma_r$ is the transverse size of the beam. For sufficiently long propagation distances, such that $|\mathbf{v}_{\perp}| d/c \gg \sigma_r$, there is a direct correspondence between the particle position at the detector and its transverse velocity, $\mathbf{x} \simeq \mathbf{v}_{\perp} d/c
\simeq  (\mathbf{p}_{\perp} / p_x)d/c$, where $\mathbf{p}_{\perp}$ is the particle transverse momentum. Figure~\ref{fig:p2p3} indicates that $|\mathbf{p}_{\perp}/p_z| \simeq 4\times10^{-4}$. Thus, after d=8\, m, $|\mathbf{x}_{\perp}|\simeq 0.32~\mathrm{cm}$, which agrees well with Fig.~\ref{fig:AtDetectorandPlasma}.

\section{Conclusions}

We show here that a virtual detector to measure the proton bunch spatial structures a few meters downstream the plasma in the AWAKE experiment provides a measurement of the transverse momentum distribution of the protons at the plasma exit rather than of the bunchlets transverse size at that location. Thus, the bunchlet radius measured at the location of the virtual detector may appear to grow along the bunch. This measurement reflects the increase of transverse momentum along the bunch, a result of the spatial growth of the plasma wakefield amplitude along the bunch. The simulations also show that the first few bunchlets may appear as a single, longer bunchlet at the detector. We note that this observation depends on the specific phase-space characteristics of the first beamlets, and also varies with the distance to the detector. In addition, transverse asymmetries in the beam momentum distribution may result in asymmetric density profiles after the plasma. Although these asymmetries could be a seed of the hosing instability, they do not necessarily lead to the growth of the hosing instability because the SMI provides an intrinsic hosing stabilisation mechanism~\cite{bib:vieira_prl}. Thus, they do not necessarily reflect the presence of the hosing instability in the plasma. Further studies to understand the observed particle distributions are ongoing.

\acknowledgements

J.V. acknowledges the support by FCT (Portugal) Grant No. SFRH/IF/01635/2015. Simulations were performed on SuperMUC at the Leibniz Research Center in Munich, Germany.

\end{document}